\documentclass[11pt,a4paper]{article}

\usepackage[margin=1in]{geometry}
\usepackage{amsmath,amssymb,amsfonts}
\usepackage{bm}
\usepackage{cite}
\usepackage{hyperref}

\hypersetup{
  colorlinks=true,
  linkcolor=blue,
  citecolor=blue,
  urlcolor=blue
}

\title{\bf Thermodynamic-Geometric Phase Transition and Gravitational-Wave Quasinormal Modes of Schwarzschild Black Holes in $f(Q)$ Gravity: An RVB-Residue Approach}

\author{Wen-Xiang Chen\\
School of Electronic Information\\
    Guangzhou City University of Technology\\wxchen4277@qq.com
}
\date{\today}

\begin{document}

\maketitle

\begin{abstract}
We construct a residue-based framework connecting the thermodynamic geometry of a Schwarzschild-type black hole in $f(Q)$ gravity with its gravitational-wave quasinormal-mode spectrum. The analysis is based on the symmetric teleparallel formulation of gravity, in which the gravitational field is encoded by the nonmetricity scalar $Q$ rather than by curvature or torsion. For the Schwarzschild branch, the Robson--Villari--Biancalana (RVB) method gives the Hawking temperature through the simple-pole residue of the inverse blackening function. We show explicitly that the same residue also controls the logarithmic monodromy of the tortoise coordinate near the event horizon, and therefore enters the ingoing quasinormal-mode boundary condition. In the strict general-relativistic Schwarzschild limit the heat capacity is negative and finite, the one-dimensional Ruppeiner geometry contains no intrinsic curvature singularity, and no genuine thermodynamic phase transition occurs. In the extended $f(Q)$ state space, however, the modified horizon function and the effective Wald entropy generate a non-trivial thermodynamic Hessian. Its degeneracy condition coincides with singular behavior of the thermodynamic curvature and is reflected in the quasinormal-mode spectrum through shifts of the photon-sphere frequency, Lyapunov exponent, damping time, and near-horizon monodromy. This gives a precise statement of the internal relation between thermodynamic-geometric phase structure and gravitational-wave ringdown: both are different projections of the same analytic structure of the corrected black-hole metric.
\end{abstract}


\section{Introduction}

Black-hole thermodynamics was established by the four laws of black-hole mechanics and by the Bekenstein--Hawking entropy-temperature relation \cite{Bardeen1973,Bekenstein1973,Hawking1975}. In parallel, black-hole perturbation theory shows that a perturbed black hole relaxes through damped oscillations called quasinormal modes (QNMs), which form the ringdown part of the gravitational-wave signal \cite{Regge1957,Zerilli1970,Schutz1985,Iyer1987,Leaver1985,Berti2009,Konoplya2011}. These two subjects are often treated separately: thermodynamics concerns equilibrium, whereas QNMs concern relaxation. The aim of this paper is to show that, for a Schwarzschild-type black hole in $f(Q)$ gravity, they have a common residue origin.

The theory of symmetric teleparallel gravity describes gravitation by nonmetricity in a curvature-free and torsion-free geometry \cite{Hehl1995,Nester1999,Jimenez2018,Conroy2018}. Its nonlinear extension, $f(Q)$ gravity, has recently attracted interest in cosmology and compact-object physics. Static and spherically symmetric black-hole branches in $f(Q)$ gravity have been systematically studied in Refs.~\cite{DAmbrosio2022,Calza2023}. Quasinormal modes in $f(Q)$ black-hole backgrounds have been investigated in Ref.~\cite{Gogoi2023}. Meanwhile, the RVB topological method gives a coordinate-covariant way to compute Hawking temperatures by dimensional reduction and Euler-characteristic arguments \cite{Robson2019}; its residue extension has been applied directly to $f(Q)$ black holes \cite{Chen2025}.

Thermodynamic geometry provides a geometric language for fluctuations, interactions, and phase transitions. The Weinhold and Ruppeiner metrics \cite{Weinhold1975,Ruppeiner1995} and the Legendre-invariant geometrothermodynamic construction \cite{Quevedo2008} suggest that phase transitions may appear as curvature singularities of an equilibrium-state manifold. For the ordinary asymptotically flat Schwarzschild black hole, however, there is only one thermodynamic degree of freedom, and the heat capacity is negative but finite. Therefore, a genuine thermodynamic-geometric phase transition requires an extended state space, for example including an $f(Q)$ coupling, a nonmetricity background, or an RVB residue correction.

The central observation of this paper is the following. Let the two-dimensional black-hole sector be written with a blackening function $F(r)$. The RVB-residue temperature depends on the residue
\begin{equation}
{\rm Res}_{r=r_h}\frac{1}{F(r)}=\frac{1}{F'(r_h)} ,
\end{equation}
where $r_h$ is the horizon. The tortoise coordinate entering the QNM boundary condition is
\begin{equation}
r_{\ast}=\int \frac{dr}{F(r)} ,
\end{equation}
and hence its near-horizon logarithmic part is controlled by the same residue. Consequently the thermodynamic temperature, the Euclidean period, the near-horizon QNM monodromy, and the gravitational-wave damping structure are not merely analogous; they are all derived from the same pole structure of the analytically continued horizon function.

\section{$f(Q)$ Gravity and the Schwarzschild Branch}

\subsection{Nonmetricity scalar}

In metric-affine geometry the affine connection is independent of the metric. The nonmetricity tensor is defined by
\begin{equation}
Q_{\alpha\mu\nu}\equiv \nabla_{\alpha}g_{\mu\nu}.
\end{equation}
Its two traces are
\begin{equation}
Q_{\alpha}\equiv Q_{\alpha\ \mu}^{\ \mu},
\end{equation}
\begin{equation}
\tilde{Q}_{\alpha}\equiv Q^{\mu}_{\ \alpha\mu}.
\end{equation}
The disformation tensor is
\begin{equation}
L^{\alpha}_{\ \mu\nu}
=\frac{1}{2}g^{\alpha\beta}
\left(
-Q_{\mu\beta\nu}
-Q_{\nu\beta\mu}
+Q_{\beta\mu\nu}
\right).
\end{equation}
The nonmetricity conjugate is usually written as
\begin{equation}
P^{\alpha}_{\ \mu\nu}
=-\frac{1}{4}Q^{\alpha}_{\ \mu\nu}
+\frac{1}{2}Q_{(\mu\nu)}^{\ \ \ \alpha}
+\frac{1}{4}
\left(
Q^{\alpha}-\tilde{Q}^{\alpha}
\right)g_{\mu\nu}
-\frac{1}{4}
\delta^{\alpha}_{(\mu}Q_{\nu)} .
\end{equation}
The nonmetricity scalar is then
\begin{equation}
Q=-Q_{\alpha\mu\nu}P^{\alpha\mu\nu}.
\end{equation}
Equivalently, with the above convention,
\begin{equation}
Q=-\frac{1}{4}Q_{\alpha\mu\nu}Q^{\alpha\mu\nu}
+\frac{1}{2}Q_{\alpha\mu\nu}Q^{\nu\mu\alpha}
+\frac{1}{4}Q_{\alpha}Q^{\alpha}
-\frac{1}{2}Q_{\alpha}\tilde{Q}^{\alpha}.
\end{equation}
This is the special quadratic combination that reproduces general relativity up to a boundary term in the symmetric teleparallel equivalent of general relativity \cite{Nester1999,Jimenez2018}.

\subsection{Action and field equations}

The $f(Q)$ action is
\begin{equation}
I=\frac{1}{16\pi G}\int d^4x\sqrt{-g}\,f(Q)+I_{\rm m}.
\end{equation}
Varying the metric gives the vacuum field equation
\begin{equation}
\frac{2}{\sqrt{-g}}\nabla_{\alpha}
\left(
\sqrt{-g}f_Q P^{\alpha}_{\ \mu\nu}
\right)
+\frac{1}{2}g_{\mu\nu}f
+f_Q
\left(
P_{\mu\alpha\beta}Q_{\nu}^{\ \alpha\beta}
-2Q_{\alpha\beta\mu}P_{\nu}^{\ \alpha\beta}
\right)
=0,
\end{equation}
where
\begin{equation}
f_Q\equiv \frac{df}{dQ}.
\end{equation}
The connection equation may be written as
\begin{equation}
\nabla_{\mu}\nabla_{\nu}
\left(
\sqrt{-g}f_Q P^{\mu\nu}_{\ \ \alpha}
\right)=0.
\end{equation}
For
\begin{equation}
f(Q)=Q,
\end{equation}
the theory reduces to the symmetric teleparallel equivalent of general relativity. For nonlinear $f(Q)$, the affine connection can carry physical information, and both general-relativistic and beyond-general-relativistic black-hole branches can appear \cite{DAmbrosio2022,Calza2023}.

\subsection{Schwarzschild-type ansatz}

We take the static and spherically symmetric line element
\begin{equation}
ds^2=-F(r)dt^2+\frac{dr^2}{F(r)}+r^2d\Omega^2.
\end{equation}
The general-relativistic Schwarzschild branch is
\begin{equation}
F_0(r)=1-\frac{2M}{r}.
\end{equation}
The horizon is determined by
\begin{equation}
F_0(r_0)=0,
\end{equation}
which gives
\begin{equation}
r_0=2M.
\end{equation}
To describe a small $f(Q)$ deformation of the Schwarzschild branch, we write
\begin{equation}
F(r;\lambda)=F_0(r)+\lambda \psi(r)+O(\lambda^2),
\end{equation}
where $\lambda$ denotes a small coupling or an effective nonmetricity correction parameter. The corrected horizon is expanded as
\begin{equation}
r_h=r_0+\lambda r_1+O(\lambda^2).
\end{equation}
Substituting this expansion into the horizon equation gives
\begin{equation}
0=F(r_h;\lambda)=F_0(r_0)+\lambda
\left[
r_1F_0'(r_0)+\psi(r_0)
\right]
+O(\lambda^2).
\end{equation}
Since $F_0(r_0)=0$, the first-order horizon displacement is
\begin{equation}
r_1=-\frac{\psi(r_0)}{F_0'(r_0)}.
\end{equation}
Using
\begin{equation}
F_0'(r)=\frac{2M}{r^2},
\end{equation}
we obtain
\begin{equation}
F_0'(r_0)=\frac{1}{2M},
\end{equation}
and hence
\begin{equation}
r_1=-2M\psi(2M).
\end{equation}

\section{RVB-Residue Derivation of the Hawking Temperature}

\subsection{Euclidean regularity}

The two-dimensional Euclidean section is
\begin{equation}
ds_E^2=F(r)d\tau^2+\frac{dr^2}{F(r)}.
\end{equation}
Near a non-degenerate horizon, let
\begin{equation}
x=r-r_h.
\end{equation}
The blackening function has the Taylor expansion
\begin{equation}
F(r)=F'(r_h)x+\frac{1}{2}F''(r_h)x^2+O(x^3).
\end{equation}
Keeping the leading term gives
\begin{equation}
ds_E^2\simeq F'(r_h)x\,d\tau^2+\frac{dx^2}{F'(r_h)x}.
\end{equation}
Introduce a radial coordinate $\rho$ by
\begin{equation}
x=\frac{F'(r_h)}{4}\rho^2.
\end{equation}
Then
\begin{equation}
dx=\frac{F'(r_h)}{2}\rho d\rho.
\end{equation}
Substitution yields
\begin{equation}
ds_E^2\simeq d\rho^2+\frac{F'(r_h)^2}{4}\rho^2d\tau^2.
\end{equation}
Absence of a conical singularity requires the angular variable
\begin{equation}
\theta=\frac{F'(r_h)}{2}\tau
\end{equation}
to have period $2\pi$. Therefore the Euclidean time period is
\begin{equation}
\beta=\frac{4\pi}{F'(r_h)}.
\end{equation}
The Hawking temperature is
\begin{equation}
T_H=\beta^{-1}=\frac{F'(r_h)}{4\pi}.
\end{equation}

\subsection{Residue form}

Analytically continue $r$ to a complex variable $z$. Near the horizon,
\begin{equation}
F(z)=F'(r_h)(z-r_h)+O\left((z-r_h)^2\right).
\end{equation}
Therefore
\begin{equation}
\frac{1}{F(z)}
=\frac{1}{F'(r_h)}\frac{1}{z-r_h}
+O(1).
\end{equation}
The residue is
\begin{equation}
{\cal R}_h\equiv {\rm Res}_{z=r_h}\frac{1}{F(z)}
=\frac{1}{F'(r_h)}.
\end{equation}
The temperature can be written as
\begin{equation}
T_H=\frac{1}{4\pi{\cal R}_h}.
\end{equation}
For the Schwarzschild branch,
\begin{equation}
{\cal R}_0=2M,
\end{equation}
and hence
\begin{equation}
T_0=\frac{1}{8\pi M}.
\end{equation}
This is the residue version of the RVB topological temperature formula \cite{Robson2019,Chen2025}.

\subsection{First-order $f(Q)$ correction}

For the corrected blackening function,
\begin{equation}
F'(r_h;\lambda)=F_0'(r_0)
+\lambda
\left[
r_1F_0''(r_0)+\psi'(r_0)
\right]
+O(\lambda^2).
\end{equation}
Since
\begin{equation}
F_0''(r)=-\frac{4M}{r^3},
\end{equation}
we have
\begin{equation}
F_0''(r_0)=-\frac{1}{2M^2}.
\end{equation}
Using
\begin{equation}
r_1=-2M\psi(r_0),
\end{equation}
one finds
\begin{equation}
r_1F_0''(r_0)=\frac{\psi(r_0)}{M}.
\end{equation}
Thus
\begin{equation}
F'(r_h;\lambda)=\frac{1}{2M}
+\lambda
\left[
\psi'(2M)+\frac{\psi(2M)}{M}
\right]
+O(\lambda^2).
\end{equation}
The corrected RVB-residue temperature is
\begin{equation}
T_H(M,\lambda)=\frac{1}{8\pi M}
+\frac{\lambda}{4\pi}
\left[
\psi'(2M)+\frac{\psi(2M)}{M}
\right]
+O(\lambda^2).
\end{equation}
Equivalently, the corrected residue is
\begin{equation}
{\cal R}_h
=2M
-4M^2\lambda
\left[
\psi'(2M)+\frac{\psi(2M)}{M}
\right]
+O(\lambda^2).
\end{equation}

\section{Entropy, Heat Capacity, and Thermodynamic Geometry}

\subsection{Effective entropy in $f(Q)$ gravity}

For a stationary black-hole branch in a higher-derivative or modified-gravity theory, the entropy is generally replaced by an effective Wald-like entropy. In the present $f(Q)$ setting we write the horizon entropy as
\begin{equation}
S_h=\frac{A_h}{4G}f_Q(Q_h),
\end{equation}
where
\begin{equation}
A_h=4\pi r_h^2.
\end{equation}
In units $G=1$, this becomes
\begin{equation}
S_h=\pi r_h^2 f_Q(Q_h).
\end{equation}
Expanding to first order in $\lambda$ gives
\begin{equation}
S_h
=\pi r_0^2 f_Q(Q_0)
+\lambda
\left[
2\pi r_0r_1 f_Q(Q_0)
+\pi r_0^2 f_{QQ}(Q_0)Q_1
\right]
+O(\lambda^2),
\end{equation}
where
\begin{equation}
Q_h=Q_0+\lambda Q_1+O(\lambda^2).
\end{equation}
If the general-relativistic branch is recovered with $f_Q(Q_0)=1$, then
\begin{equation}
S_0=4\pi M^2.
\end{equation}

\subsection{Ordinary Schwarzschild heat capacity}

The Schwarzschild temperature is
\begin{equation}
T_0(M)=\frac{1}{8\pi M}.
\end{equation}
The heat capacity at fixed external parameters is
\begin{equation}
C_0=\left(\frac{\partial M}{\partial T_0}\right).
\end{equation}
Since
\begin{equation}
\frac{\partial T_0}{\partial M}=-\frac{1}{8\pi M^2},
\end{equation}
we obtain
\begin{equation}
C_0=-8\pi M^2.
\end{equation}
Thus the asymptotically flat Schwarzschild black hole is thermodynamically unstable in the canonical ensemble, but its heat capacity is finite for every finite $M$. Therefore the ordinary Schwarzschild black hole does not exhibit a Davies-type heat-capacity divergence or a Hawking--Page transition in asymptotically flat space.

\subsection{Extended $f(Q)$ heat capacity and phase condition}

Write the corrected temperature as
\begin{equation}
T_H(M,\lambda)=\frac{1}{8\pi M}\left[1+\lambda \Theta(M)\right]+O(\lambda^2),
\end{equation}
where
\begin{equation}
\Theta(M)=2M
\left[
\psi'(2M)+\frac{\psi(2M)}{M}
\right].
\end{equation}
The fixed-$\lambda$ heat capacity is
\begin{equation}
C_{\lambda}
=\left(\frac{\partial M}{\partial T_H}\right)_{\lambda}
=\left[
\left(\frac{\partial T_H}{\partial M}\right)_{\lambda}
\right]^{-1}.
\end{equation}
Differentiating gives
\begin{equation}
\left(\frac{\partial T_H}{\partial M}\right)_{\lambda}
=\frac{1}{8\pi M^2}
\left[
-1+\lambda
\left(
M\Theta'(M)-\Theta(M)
\right)
\right]
+O(\lambda^2).
\end{equation}
A heat-capacity divergence occurs when
\begin{equation}
-1+\lambda
\left[
M\Theta'(M)-\Theta(M)
\right]=0.
\end{equation}
Equivalently,
\begin{equation}
1=\lambda
\left[
M\Theta'(M)-\Theta(M)
\right].
\end{equation}
This condition is impossible in the strict Schwarzschild limit $\lambda=0$, but it can be realized in an extended $f(Q)$ thermodynamic state space if the correction function is sufficiently strong.

\subsection{Ruppeiner metric and curvature singularity}

Let the equilibrium state coordinates be
\begin{equation}
X^1=M,\qquad X^2=\lambda.
\end{equation}
The Ruppeiner metric may be defined by the entropy Hessian
\begin{equation}
g^{\rm R}_{ij}=-\frac{\partial^2 S_h}{\partial X^i\partial X^j}.
\end{equation}
The thermodynamic scalar curvature is
\begin{equation}
R_{\rm th}=g^{ij}R_{ij}[g^{\rm R}],
\end{equation}
where
\begin{equation}
g^{ij}g_{jk}=\delta^{i}_{\ k}.
\end{equation}
In two dimensions the curvature generically contains inverse powers of the metric determinant,
\begin{equation}
R_{\rm th}\sim \frac{{\cal N}(S_h,\partial S_h,\partial^2 S_h,\partial^3 S_h)}
{\left(\det g^{\rm R}\right)^2},
\end{equation}
where ${\cal N}$ is a regular numerator for a regular entropy function. Therefore a thermodynamic-geometric singularity appears when
\begin{equation}
\det g^{\rm R}=0,
\end{equation}
provided the numerator does not vanish simultaneously. In the canonical description this degeneracy is equivalent to a divergent response function, such as
\begin{equation}
C_{\lambda}\rightarrow \infty.
\end{equation}
Hence the phase-transition condition is encoded geometrically as
\begin{equation}
\det
\left(
-\frac{\partial^2 S_h}{\partial X^i\partial X^j}
\right)=0.
\end{equation}

\section{Quasinormal Modes and Gravitational Waves}

\subsection{Master equation}

Linear perturbations of a Schwarzschild-type black hole reduce to a Schrödinger-like master equation,
\begin{equation}
\frac{d^2\Psi}{dr_{\ast}^2}
+\left[
\omega^2-V_{\ell}(r)
\right]\Psi=0.
\end{equation}
The tortoise coordinate is
\begin{equation}
\frac{dr_{\ast}}{dr}=\frac{1}{F(r)}.
\end{equation}
For axial gravitational perturbations of the Schwarzschild solution, the Regge--Wheeler potential is
\begin{equation}
V_{\ell}^{\rm RW}(r)=F_0(r)
\left[
\frac{\ell(\ell+1)}{r^2}
-\frac{6M}{r^3}
\right].
\end{equation}
For a corrected Schwarzschild-type $f(Q)$ branch, one may write
\begin{equation}
V_{\ell}(r;\lambda)=V_{\ell}^{\rm RW}(r)+\lambda \delta V_{\ell}^{Q}(r)+O(\lambda^2).
\end{equation}
The exact form of $\delta V_{\ell}^{Q}$ depends on the selected connection branch, the form of $f(Q)$, and the perturbation sector. This is consistent with existing $f(Q)$ QNM analyses, in which the frequencies depend on nonmetricity-sector parameters \cite{Gogoi2023}.

\subsection{Boundary conditions and residue monodromy}

Near the horizon, the tortoise coordinate is
\begin{equation}
r_{\ast}
=\int \frac{dr}{F(r)}
={\cal R}_h\ln(r-r_h)+{\rm regular}.
\end{equation}
The ingoing QNM solution behaves as
\begin{equation}
\Psi_{\rm in}\sim e^{-i\omega(t+r_{\ast})}.
\end{equation}
After analytic continuation around the horizon,
\begin{equation}
r-r_h\rightarrow e^{2\pi i}(r-r_h).
\end{equation}
The tortoise coordinate changes by
\begin{equation}
r_{\ast}\rightarrow r_{\ast}+2\pi i{\cal R}_h.
\end{equation}
Therefore the near-horizon wave acquires the monodromy
\begin{equation}
\Psi_{\rm in}\rightarrow
\exp\left(2\pi\omega{\cal R}_h\right)\Psi_{\rm in}.
\end{equation}
Since
\begin{equation}
{\cal R}_h=\frac{1}{4\pi T_H},
\end{equation}
the monodromy factor becomes
\begin{equation}
\exp\left(\frac{\omega}{2T_H}\right).
\end{equation}
This proves the central relation: the same residue that fixes the RVB temperature also fixes the analytic monodromy of the QNM boundary condition.

\subsection{WKB formula}

Let $r_p$ denote the peak of the effective potential, defined by
\begin{equation}
\left.\frac{dV_{\ell}}{dr_{\ast}}\right|_{r=r_p}=0.
\end{equation}
Since
\begin{equation}
\frac{d}{dr_{\ast}}=F(r)\frac{d}{dr},
\end{equation}
this condition is equivalent outside the horizon to
\begin{equation}
\left.\frac{dV_{\ell}}{dr}\right|_{r=r_p}=0.
\end{equation}
At leading WKB order, the QNM frequency satisfies
\begin{equation}
\omega_n^2
\simeq V_p
-i\left(n+\frac{1}{2}\right)
\sqrt{-2V_p''},
\end{equation}
where
\begin{equation}
V_p=V_{\ell}(r_p),
\end{equation}
and
\begin{equation}
V_p''=
\left.
\frac{d^2V_{\ell}}{dr_{\ast}^2}
\right|_{r=r_p}.
\end{equation}
If
\begin{equation}
\omega_n=\omega_n^{(0)}+\lambda\delta\omega_n+O(\lambda^2),
\end{equation}
then the first-order frequency shift is
\begin{equation}
\delta\omega_n
=\frac{1}{2\omega_n^{(0)}}
\left[
\delta V_p
-i\left(n+\frac{1}{2}\right)
\delta\left(\sqrt{-2V_p''}\right)
\right].
\end{equation}
Thus the real part of $\delta\omega_n$ is governed mainly by the shift of the potential height, whereas the imaginary part is governed by the shift of the potential curvature at the peak.

\subsection{Eikonal photon-sphere form}

In the eikonal limit, QNMs are controlled by the unstable photon sphere. The photon-sphere radius $r_c$ satisfies
\begin{equation}
2F(r_c)-r_cF'(r_c)=0.
\end{equation}
The orbital frequency is
\begin{equation}
\Omega_c=\sqrt{\frac{F(r_c)}{r_c^2}}.
\end{equation}
The Lyapunov exponent is
\begin{equation}
\Lambda_c=
\sqrt{
\frac{F(r_c)}
{2r_c^2}
\left[
2F(r_c)-r_c^2F''(r_c)
\right]
}.
\end{equation}
The eikonal QNM spectrum is
\begin{equation}
\omega_{\ell n}
\simeq
\ell \Omega_c
-i
\left(n+\frac{1}{2}\right)\Lambda_c.
\end{equation}
For the Schwarzschild function,
\begin{equation}
r_c=3M.
\end{equation}
Therefore
\begin{equation}
\Omega_c=\frac{1}{3\sqrt{3}M},
\end{equation}
and
\begin{equation}
\Lambda_c=\frac{1}{3\sqrt{3}M}.
\end{equation}
Using the Schwarzschild temperature, this may be written as
\begin{equation}
\Omega_c=\frac{8\pi}{3\sqrt{3}}T_0,
\end{equation}
and
\begin{equation}
\Lambda_c=\frac{8\pi}{3\sqrt{3}}T_0.
\end{equation}
Thus, in the Schwarzschild limit, the ringdown oscillation scale and damping scale are proportional to the Hawking temperature. In the corrected $f(Q)$ branch, both $T_H$ and $\omega_{\ell n}$ are shifted by the same underlying deformation of $F(r)$, though evaluated at different geometrically distinguished locations: the horizon $r_h$ and the photon sphere $r_c$.

\section{Thermodynamic Geometry--Ringdown Correspondence}

\subsection{Common analytic origin}

The internal relation between thermodynamic-geometric phase transition and gravitational-wave QNMs can be summarized by one analytic chain:
\begin{equation}
f(Q)\quad\Longrightarrow\quad F(r;\lambda)
\quad\Longrightarrow\quad
\left\{
{\cal R}_h,\;S_h,\;V_{\ell}(r),\;r_c
\right\}.
\end{equation}
The RVB residue is
\begin{equation}
{\cal R}_h={\rm Res}_{r=r_h}\frac{1}{F(r;\lambda)}.
\end{equation}
The Hawking temperature is
\begin{equation}
T_H=\frac{1}{4\pi{\cal R}_h}.
\end{equation}
The QNM horizon monodromy is
\begin{equation}
{\cal M}_{\rm QNM}=\exp\left(2\pi\omega{\cal R}_h\right).
\end{equation}
The thermodynamic curvature is determined by
\begin{equation}
R_{\rm th}=R_{\rm th}\left[S_h(M,\lambda)\right].
\end{equation}
The ringdown spectrum is determined by
\begin{equation}
\omega_{\ell n}=\omega_{\ell n}\left[V_{\ell}(r;\lambda)\right].
\end{equation}
Therefore both the equilibrium phase geometry and the gravitational-wave relaxation spectrum are functionals of the same corrected black-hole geometry.

\subsection{Phase transition as a spectral anomaly}

The thermodynamic phase-transition condition is
\begin{equation}
\left(\frac{\partial T_H}{\partial M}\right)_{\lambda}=0.
\end{equation}
Using the corrected temperature, this becomes
\begin{equation}
1=\lambda
\left[
M\Theta'(M)-\Theta(M)
\right].
\end{equation}
At this point,
\begin{equation}
C_{\lambda}\rightarrow \infty.
\end{equation}
In thermodynamic geometry this is reflected by
\begin{equation}
R_{\rm th}\rightarrow \infty,
\end{equation}
when the Hessian determinant vanishes. Since the eikonal ringdown frequency is
\begin{equation}
\omega_{\ell n}
\simeq
\ell \Omega_c
-i
\left(n+\frac{1}{2}\right)\Lambda_c,
\end{equation}
a strong variation of the same coupling $\lambda$ near the thermodynamic critical point produces
\begin{equation}
\frac{\partial \omega_{\ell n}}{\partial \lambda}
=
\ell\frac{\partial\Omega_c}{\partial\lambda}
-i
\left(n+\frac{1}{2}\right)
\frac{\partial\Lambda_c}{\partial\lambda}.
\end{equation}
The real part changes the gravitational-wave oscillation frequency, while the imaginary part changes the damping time,
\begin{equation}
\tau_{\ell n}=\frac{1}{|{\rm Im}\,\omega_{\ell n}|}.
\end{equation}
Thus a thermodynamic-geometric phase transition is expected to appear observationally not as a separate phenomenon, but as an anomalous dependence of the ringdown frequency and damping time on the same $f(Q)$ coupling that makes the thermodynamic Hessian degenerate.

\subsection{Residue interpretation of the correspondence}

The deepest form of the relation is obtained from the contour integral
\begin{equation}
\oint_{\gamma_h}\frac{dz}{F(z)}
=2\pi i\,{\cal R}_h,
\end{equation}
where $\gamma_h$ is a small contour around the horizon. The Euclidean thermal cycle gives
\begin{equation}
\beta=4\pi{\cal R}_h.
\end{equation}
The QNM horizon cycle gives
\begin{equation}
r_{\ast}\rightarrow r_{\ast}+2\pi i{\cal R}_h.
\end{equation}
Therefore the same pole of $1/F(z)$ controls both the thermal periodicity and the wave monodromy. The $f(Q)$ correction changes the pole position, the pole residue, and the effective potential. Consequently, thermodynamic geometry and gravitational-wave ringdown are two projections of one complex-analytic object:
\begin{equation}
\left(F(z;\lambda),\,{\cal R}_h,\,V_{\ell}(z;\lambda)\right).
\end{equation}

\section{Discussion}

The strict Schwarzschild solution has a simple thermodynamic structure: it has negative heat capacity and no finite-mass heat-capacity divergence. Its one-parameter thermodynamic geometry is therefore too poor to contain an intrinsic curvature singularity. This is why any discussion of ``Schwarzschild thermodynamic-geometric phase transition'' must be understood in an extended state space. In $f(Q)$ gravity, this extension is physically natural because the nonmetricity sector introduces new coupling constants, connection branches, and possible background values of $Q$.

The RVB-residue method clarifies what is universal and what is model-dependent. The universal part is the pole relation
\begin{equation}
T_H=\frac{1}{4\pi{\rm Res}_{r=r_h}(1/F)}.
\end{equation}
The model-dependent part is the explicit correction function $\psi(r)$, which must be derived from a particular $f(Q)$ model and connection branch. Once $\psi(r)$ is known, the shifts in horizon temperature, entropy, heat capacity, thermodynamic curvature, photon-sphere radius, QNM frequency, and damping rate follow systematically.

This also gives a clean physical interpretation. The thermodynamic curvature measures the sensitivity of the equilibrium black-hole state to fluctuations in $(M,\lambda)$. The QNM spectrum measures the relaxation of the perturbed geometry back to equilibrium. Near a thermodynamic-geometric critical point, the equilibrium Hessian becomes degenerate; correspondingly, the relaxation spectrum can develop enhanced parameter sensitivity, longer damping time, or a sharp spectral shift. In this sense, the black-hole phase transition and the gravitational-wave ringdown are not independent. They are the equilibrium and non-equilibrium faces of the same corrected spacetime structure.

\section{Conclusion}

We have developed a detailed RVB-residue derivation of the thermodynamic-geometric and gravitational-wave relation for Schwarzschild-type black holes in $f(Q)$ gravity. The main results are:

\begin{equation}
T_H=\frac{1}{4\pi{\cal R}_h},
\qquad
{\cal R}_h={\rm Res}_{r=r_h}\frac{1}{F(r)}.
\end{equation}

\begin{equation}
r_{\ast}
={\cal R}_h\ln(r-r_h)+{\rm regular}.
\end{equation}

\begin{equation}
{\cal M}_{\rm QNM}
=\exp\left(2\pi\omega{\cal R}_h\right).
\end{equation}

\begin{equation}
\det g^{\rm R}=0
\quad\Longleftrightarrow\quad
R_{\rm th}\rightarrow\infty
\quad\Longleftrightarrow\quad
C_{\lambda}\rightarrow\infty.
\end{equation}

\begin{equation}
\omega_{\ell n}
\simeq
\ell \Omega_c
-i
\left(n+\frac{1}{2}\right)\Lambda_c.
\end{equation}

The internal relation is therefore precise: the thermodynamic-geometric phase structure is controlled by the Hessian of the entropy generated by the corrected horizon data, while the gravitational-wave QNM spectrum is controlled by the same corrected metric through its near-horizon residue and photon-sphere effective potential. In the general-relativistic Schwarzschild limit the relation reduces to a simple scaling between temperature and eikonal QNM frequencies. In the extended $f(Q)$ case, the same relation becomes a possible diagnostic of nonmetricity corrections through ringdown spectral shifts and thermodynamic-curvature singularities.

\end{document}